\documentclass[]{svjour3}
\usepackage{amsmath,amssymb}
\usepackage{graphicx}
\usepackage[colorlinks=true, citecolor=blue, urlcolor=blue ]{hyperref}
\newcommand{\bq}{\begin{equation}}
\newcommand{\eq}{\end{equation}}
\newcommand{\bqs}{\begin{equation*}}
\newcommand{\eqs}{\end{equation*}}
\newcommand{\ba}{\begin{array}}
\newcommand{\ea}{\end{array}}
\newcommand{\bas}{\begin{array*}}
\newcommand{\eas}{\end{array*}}
\newcommand{\bqa}{\begin{eqnarray}}
\newcommand{\eqa}{\end{eqnarray}}
\newcommand{\bqas}{\begin{eqnarray*}}
\newcommand{\eqas}{\end{eqnarray*}}

\newcommand{\etal}{\textit{et al.}}

\newcommand{\ran}{\rangle}
\newcommand{\lan}{\langle}

\newcommand{\D}{\mathcal{D}}
\newcommand{\N}{\mathcal{M}}

\begin{document}
\title{Geometric discord and Measurement-induced nonlocality for well known bound entangled states}
\titlerunning{GD and MIN for well known bound entangled states}
\author{Swapan Rana \and Preeti Parashar}
\institute{Physics and Applied Mathematics
Unit, Indian Statistical Institute,\\ 203 B T Road, Kolkata 700
108, India\\
\email{swapanqic@gmail.com, parashar@isical.ac.in}}

\maketitle

\begin{abstract}
We employ geometric discord and measurement induced nonlocality to quantify non classical correlations of some well-known bipartite bound entangled states, namely the two families of Horodecki's ($2\otimes 4$, $3\otimes 3$ and $4\otimes 4$ dimensional) bound entangled states and that of Bennett \etal's in $3\otimes 3$ dimension. In most of the cases our results are analytic and both the measures attain relatively small value. The amount of quantumness in the $4\otimes 4$ bound entangled state of Benatti \etal and the $2\otimes 8$ state having the same matrix representation (in computational basis) is same. Coincidently, the $2m\otimes 2m$ Werner and isotropic states also exhibit the same property, when seen as $2\otimes 2m^2$ dimensional states.
\end{abstract}

\section{Introduction} Entanglement is the most mysterious feature of quantum mechanics since its early days. In the last two decades, with the emergence of quantum information theory, entanglement has been recognized as a resource. There are two kinds of entanglement, the \emph{free} entanglement which can be used for various information processing tasks like teleportation, super dense coding etc., and the \emph{bound} entanglement which needs a finite amount of free entanglement for its creation, but once created, no free entanglement can be extracted (distilled) from it \cite{MPRHorodeckiPRL98}. The latter one is the most peculiar in the sense that in contrast to its counterpart, it can not be used directly (alone)  but when supplied with other resources, can enhance the efficiency of some tasks \cite{PHorodeckiandRHorodeckiQIC01}. It is indeed very difficult to detect and even more difficult to quantify bound entanglement. To the best of our knowledge, only two works have quantified entanglement in some well known classes of bound entangled states, one pertaining to bipartite states \cite{SGharibianetalQIC09}, and the other to multipartite states \cite{TCWeietalPRA04}.

Until a few years ago, entanglement was the only \emph{hallmark} of \emph{quantumness}, i.e., absence of entanglement and \emph{classicality} were the same notions. With the work of Ollivier and Zurek \cite{OllivierandZurekPRL01}, the notion of quantumness has been extended far beyond entanglement and several measures of nonclassicality have been proposed. Of them, the two interesting ones are the geometric discord (GD) and the measurement-induced nonlocality (MIN). In this work we evaluate these measures for some well-known bound entangled states. We note that the negativity \cite{VidalWernerPRA02}, the most common measure of mixed state entanglement, can not quantify quantumness of these states.

The concept of GD was introduced by Daki\'{c} \etal \cite{DakicetalPRL10} and for an $m\otimes n$ ($m\le n$) state $\rho$ it is defined as (normalized such that the maximum is $1$) \bq\label{defgd} \D(\rho)=\frac{m}{m-1}\min_{\chi\in\Omega_0}\|\rho-\chi\|^2\eq where $\Omega_0$ is the set of zero-discord states (i.e., \emph{classical-quantum states}, given by $\sum p_k|\psi_k\ran\lan \psi_k|\otimes\rho_k$) and $\|A\|^2=$ Tr$(A^\dagger A)$ is the Frobenius or Hilbert-Schmidt norm. Later on Luo and Fu \cite{LuoFuPRA10} gave an alternative description of GD via a minimization over all possible von Neumann measurements $\Pi^A=\{\Pi_k^A\}$ on $\rho^A$, \bq\label{drhopi}\D(\rho)=\frac{m}{m-1}\min_{\Pi^A}\|\rho-\Pi^A(\rho)\|^2\eq where $\Pi^A(\rho):=\sum_k (\Pi_k^A\otimes I^B)\rho(\Pi_k^A\otimes I^B)$. Very recently GD has been demonstrated to be a better resource than entanglement for remote state preparation \cite{DakicetalAR12}.

Let $\rho$ has the following Bloch form
\bq\label{blochrho}\rho=\frac{1}{mn}\left[I_m\otimes I_n+\mathbf{x}^t\mathbf{\mu}\otimes I_n+I_m\otimes\mathbf{y}^t\mathbf{\nu}+\sum T_{ij}\mu_i\otimes \nu_j\right]\eq where $\mathbf{\mu}=(\mu_1,\mu_2,\ldots,\mu_{m^2-1})^t$ with $\mu_i$ being the generators of $SU(m)$ and similarly for $\nu$ \cite{BertlmannKrammerJPA08}. Then a lower bound on  GD is given by \cite{RanaParasharPRA12,HassanetalPRA12}
\bq\label{lowerboundgd} \D(\rho)\ge\frac{2}{m(m-1)n}\left[\mathbf\|{x}\|^2+\frac{2}{n}\|T\|^2-\sum_{k=1}^{m-1}\lambda_k^{\downarrow}(G)\right]\eq where $\lambda_{k}^{\downarrow}(G)$ are the eigenvalues of $G:=xx^t+\frac{2}{n}TT^t$ sorted in non-increasing order. The bound is saturated for all $2\otimes n$ states and in this case the optimal measurements are given by $ \Pi_{1,2}=\frac{1}{2}(\mathbf{I}\pm\mathbf{e}.\mu)$, $\mathbf{e}$ being the eigenvector corresponding to the eigenvalue  $\lambda_1^{\downarrow}$. Similarly, for $3\otimes3$ systems, if the following operators \bq\label{optimalgdporj}\Pi_{1,2}=\frac{1}{3}I+\frac{1}{2}\left(\pm \mathbf{e}_1+\frac{1}{\sqrt{3}}\mathbf{e}_2\right)\mu,\quad\Pi_3=I-\Pi_1-\Pi_2\eq are legitimate orthonormal projectors, they constitute the optimal (not necessarily unique) von Neumann measurement and equality holds in Eq. (\ref{lowerboundgd}).

The other measure of quantumness (or rather non-locality), namely the MIN is somewhat dual to GD and is defined as (normalized)\footnote{To avoid notational conflict with a well known entanglement measure, negativity \cite{VidalWernerPRA02}, we use $\N$ instead of $\mathcal{N}$.} \cite{LuoFuPRL11} \bq\label{defmin}\N(\rho)=\frac{m}{m-1}\max_{\Pi^A}\|\rho-\Pi^A(\rho)\|^2\eq where the maximization is over all von Neumann measurements $\Pi^A$ which \emph{do not disturb} $\rho^A$ locally, i.e., $\sum_k\Pi^A_k\rho^A\Pi^A_k=\rho^A$. The MIN can be non-zero even for a zero-discord state, although an operational interpretation is not yet known . If $\rho^A$ is non-degenerate, its eigen-projectors correspond to the unique optimal von Neumann measurement, otherwise we have the following upperbound on MIN \cite{LuoFuPRL11,RanaParasharPRA12} \bq\label{upperboundmin} \N(\rho)\le \frac{4}{m(m-1)n^2}\sum_{k=1}^{m^2-m}\lambda_k^{\downarrow} (TT^t)\eq Similar to the case of GD, this bound is saturated for all $2\otimes n$ states. The two bounds given by Eq. (\ref{lowerboundgd}) and Eq. (\ref{upperboundmin}) are the main tools for our analytical study. In other situations, we shall resort to numerical techniques.

\section{Horodecki's bound entagled states}
\subsection{P. Horodecki's $2\otimes4$ bound entangled state \cite{HorodeckiPLA97}} The first bound entangled state in the literature (and also in the least possible dimension) is given by
\bq\label{phbes24} \rho_a=\frac{7a}{7a+1}\rho_{\mbox{ent}}+\frac{1}{7a+1}|\phi\ran\lan\phi|,\quad a\in[0,1]\eq where
\bqa |\phi\ran&=& |1\ran\otimes\left( \sqrt{\frac{1+a}{2}}|0\ran+\sqrt{\frac{1-a}{2}}|2\ran\right),\nonumber\\
\rho_{\mbox{ent}}&=&\frac{2}{7}\sum_{i=1}^3|\psi_i\ran\lan\psi_i|+\frac{1}{7}|03\ran\lan03|,\nonumber\\
|\psi_i\ran&=&\frac{1}{\sqrt{2}}(|0i-1\ran+|1i\ran),\quad i=1,2,3.\nonumber\eqa

 The state $\rho_a$ remains PPT throughout $a\in[0,1]$ and for $0<a<1$, it is bound entangled. Expressed in the Bloch form, we have $x=(0,0,\frac{a-1}{7a+1})^t$ and the eigenvalues of $G$ are $\frac{24 a^2}{(1+7 a)^2}$ (multiplicity $2$) and $\frac{4 (1-a)}{(1+7 a)^2}$. Since for $m=2$, equality holds in the formula (\ref{lowerboundgd}), we have
 \bq\label{gdph24}\D(\rho_a)=\left\{\begin{array}{ll}
\frac{12 a^2}{(1+7 a)^2}, & \text{ if } a\in[0, \frac{1}{3}]\\
\frac{1+ a (6 a-1)}{(1+7 a)^2}, & \text{ if } a\in[\frac{1}{3},1]
\end{array}\right.\eq
The corresponding optimal von Neumann projectors are given by\bqs\Pi_{1,2}=\left\{\begin{array}{ll}
\frac{1}{2}(\mathbf{I}\pm\sigma_z), & \text{ if } a\in[0, \frac{1}{3}]\\
\frac{1}{2}(\mathbf{I}\pm\sigma_x)\text{ or }\frac{1}{2}(\mathbf{I}\pm\sigma_y), & \text{ if } a\in[\frac{1}{3},1]
\end{array}\right.\eqs

Let us now evaluate MIN. The eigenvalues of $\rho^A$ are $\frac{4 a}{1+7 a}$ and $\frac{1+3 a}{1+7 a}$. So $\rho_a$ is non-degenerate unless $a=1$. In the non-degenerate case, the optimal von Neumann measurement (which corresponds to the eigen-projectors of $\rho^A$) is given by $\{|0\ran\lan0|,|1\ran\lan1|\}$. Therefore, the MIN is exactly the same as GD in the first case in Eq. (\ref{gdph24}). In the degenerate case, since equality holds in Eq. (\ref{upperboundmin}), we have $\N=3/16$. This matches with the MIN of non-degenerate case, if we put $a=1$. Thus we have
\bq\label{finalminph24} \N(\rho_a)=\frac{12 a^2}{(1+7 a)^2}\eq and the optimal measurement is $\{|0\ran\lan0|,|1\ran\lan1|\}$.

We note that for $a=0$, $\rho_a$ becomes a product state and has a vanishing $\D$ and $\N$. However, for $a=1$, though separable, $\rho_a$ has non-vanishing $\D$ and $\N$.

We also note that the maximum value of $\D$ is $3/25$ (occurs at $a=1/3$) and the maximum value of $\N$ is $3/16$ (occurs at $a=1$) which are quite small compared to the maximum possible value 1.

\subsection{Horodeckis' $3\otimes3$ bound entangled state \cite{AllHorodecki01,PKrammerDT}} The state is given by
\bq\label{allhbes} \rho_\beta=\frac{2}{7}|\Phi\ran\lan\Phi|+\frac{\beta}{7}\sigma_+ +\frac{5-\beta}{7}\sigma_-,\quad \beta\in[0,5]\eq where
\bqas |\Phi\ran&=& \frac{1}{\sqrt{3}}\sum_{k=0}^2|kk\ran\\
\sigma_+&=&\frac{1}{3}\left( |01\ran\lan01|+|12\ran\lan12|+|20\ran\lan20|\right)\\
\sigma_-&=&\frac{1}{3}\left( |10\ran\lan10|+|21\ran\lan21|+|02\ran\lan02|\right)\eqas

This is an interesting class as it contains all kinds of states depending on the allowed range of the defining parameter $\beta$. Surprisingly, it turns out that this class is also a good example for demonstrating the duality between $\D$ and $\N$.

For this state explicit calculation shows that x=0 and the eigenvalues of G are $\frac{6}{49}$ (multiplicity $6$) and $\frac{3}{98} (19-15 \beta +3 \beta ^2)$ (multiplicity $2$). Accordingly, we have the following bound
\bq\label{allhboundgd}\D(\rho_{\beta})\ge\left\{\begin{array}{ll}
\frac{1}{49} \left(9-5 \beta +\beta ^2\right), & \text{ if } \beta\in[\frac{5-\sqrt{5}}{2}, \frac{5+\sqrt{5}}{2}]\\
\frac{4}{49}, & \text{ otherwise }
\end{array}\right.\eq

In the first case, the operators in (\ref{optimalgdporj}) fail to be orthonormal projectors and so we can not guarantee the equality. Indeed, the best we achieved (which is not by systematic procedure, but by hit and trial) is the value $\frac{1}{294} (49-25 \beta +5 \beta ^2)$, corresponding to the measurement $\Pi_k=|p_k\ran\lan p_k|,k=1,2$, with $p_1=(1,1,1)/\sqrt{3},p_2=(1,1,-2)/\sqrt{6}$. Thus, for $\beta\in[\frac{5-\sqrt{5}}{2}, \frac{5+\sqrt{5}}{2}]$, we have \bq\label{allhboundgdboth}\frac{1}{49} \left(9-5 \beta +\beta ^2\right)\le\D(\rho_{\beta})\le\frac{1}{294}\left(49-25 \beta +5 \beta ^2\right)\eq In the second case, however, the operators in (\ref{optimalgdporj}) are orthonormal projectors ($\Pi_k=|k\ran\lan k|$) and hence we have the exact value of GD.

Now to evaluate MIN, we note that $\rho^A$ is degenerate (since $x=0$). So using Eq. (\ref{upperboundmin}), we have the following upperbound on MIN
\bq\label{allhboundmin}\N(\rho_{\beta})\le\left\{\begin{array}{ll}
\frac{4}{49}, & \text{ if } \beta\in[\frac{5-\sqrt{5}}{2}, \frac{5+\sqrt{5}}{2}]\\
\frac{1}{49} \left(9-5 \beta +\beta ^2\right), & \text{ otherwise }
\end{array}\right.\eq

\begin{figure}[ht]
  \includegraphics[height=4.5cm]{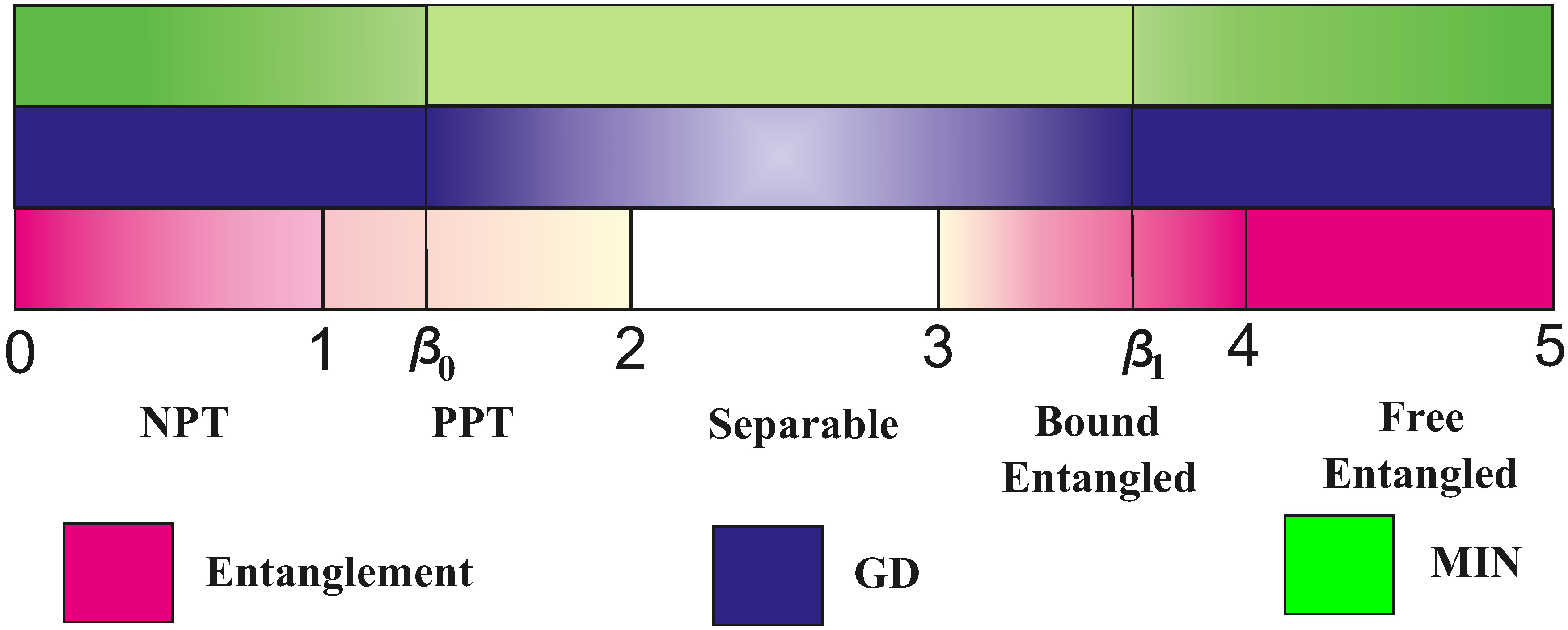}
  \caption{(Color online) Schematic presentation of quantumness of the state $\rho_{\beta}$ given by Eq. (\ref{allhbes}). For $0\le\beta<1$, the state is NPT and hence entangled (but it is not known whether free or bound), for $1\le\beta\le 4$ the state is PPT (but unknown whether separable or entangled in $1\le\beta<2$), for $2\le\beta\le 3$ separable, for $3<\beta\le 4$ bound entangled and for $4<\beta\le 5$ it is free entangled \cite{PKrammerDT}. The state has varying GD within $\beta_0<\beta<\beta_1$ ($\D$ decreases from $\beta_0$ to $5/2$ and then increases symmetrically) but constant outside. Contrary to GD, MIN is fixed within $\beta_0<\beta<\beta_1$ and varies outside ($\N$ decreases from $0$ to $\beta_0$, then remains fixed up to $\beta_1$ and  increases symmetrically from $\beta_1$ to $5$. This demonstrates duality between GD and MIN. Note that both $\D$ and $\N$ are always strictly positive irrespective of entanglement. The points $\beta_{0,1}$ are ($5\mp\sqrt{5})/2$.}\label{allhfig}
\end{figure}

Opposite to the case of GD, here equality holds in the first case while for the second case we have \bq\label{allhboundminboth}\frac{1}{294}\left(49-25 \beta +5 \beta ^2\right)\le\N(\rho_{\beta})\le\frac{1}{49} \left(9-5 \beta +\beta ^2\right)\eq We note that Eqs. (\ref{allhboundgd})-(\ref{allhboundminboth}) clearly demonstrate the duality between GD and MIN. We also note that $\D_{\max}=4/49, \D_{\min}\ge 11/196$ and $\N_{\min}=4/49, \N_{\max}\le 1/6$. Thus both GD and MIN are small, but always strictly positive, irrespective of entanglement. A schematic comparison between entanglement, GD and MIN has been illustrated in FIG-\ref{allhfig}.

\subsection{Horodecki's $4\otimes 4$ bound entangled state having a positive distillable secure key rate \cite{HorodeckietalIEEE08}}
This state is particularly interesting, as it is the simplest example of a bound entangled state featuring privacy via distillable secret key.
The state is formally given by \bq \rho_h= \sum_{i=0}^3q_i|\psi_i\ran\lan\psi_i|_{AB}\otimes \rho^{(i)}_{A'B'}\eq where \bqas \rho^{(0)}&=&\frac{1}{2}\left(|00\ran\lan 00|+|\psi_2\ran\lan\psi_2|\right),\\\rho^{(1)}&=&\frac{1}{2}\left(|11\ran\lan 11|+|\psi_3\ran\lan\psi_3|\right),\\\rho^{(2,3)}&=&|\chi_{\pm}\ran\lan\chi_{\pm}|,\eqas
with $|\psi_{0,1}\ran=(|00\ran\pm|11\ran)/\sqrt{2}$, $|\psi_{2,3}\ran=(|01\ran\pm|10\ran)/\sqrt{2}$, $|\chi_{\pm}\ran=(|00\ran\pm|\psi_0\ran)/\sqrt{2\pm\sqrt{2}}$; the mixing parameters $\{q_i\}_{i=1}^3$ are given by $\{p/2,p/2,(1-p)/2,(1-p)/2\}$ with
$p=\sqrt{2}/(1+\sqrt{2})$. 

For calculational simplicity, we will first write this state in the computational basis (omitting the zero entries) as 
\bq\rho =\left(
\begin{array}{cccc|cccc|cccc|cccc}
 s &  &  &  &  &  &  &  &  &  & s & &  &  &  & \\
  & s &  &  &  &  &  &  &  &  &  &  &  &  & s &  \\
  &  & t &  &  &  &  &  & s & &  &  &  & s &  &  \\
  &  &  &  &  &  &  &  &  &  &  & &  &  &  &  \\\hline
  &  &  &  & s &  &  &  &  &  &  & s &  &  &  &  \\
  &  &  &  &  & s &  &  &  &  &  &  &  &  &  & -s \\
  &  &  &  &  &  &  &  &  &  &  &  &  &  &  &  \\
  &  &  &  &  &  &  & t & s &  &  &  &  & -s &  & \\\hline
  &  & s &  &  &  &  & s & t &  &  &  &  &  &  &  \\
  &  &  &  &  &  &  &  &  &  &  &  &  &  &  &  \\
 s &  &  &  &  &  &  &  &  &  & s &  &  &  &  &  \\
  &  &  &  & s &  &  &  &  &  &  & s &  &  &  &  \\\hline
  &  &  &  &  &  &  &  &  &  &  &  &  &  &  &  \\
  &  & s &  &  &  &  & -s &  &  &  &  &  & t &  &  \\
  & s &  &  &  &  &  &  &  &  &  &  &  &  & s &  \\
  &  &  &  &  & -s &  &  &  &  &  &  &  &  &  & s
\end{array}
\right)\eq
where $s=p/8=\sqrt{2}/8(1+\sqrt{2}), t=p/4\sqrt{2}=1/4(1+\sqrt{2})$. Now it is easy to calculate the bounds (note that $x=0$) and we have 
\bqa\label{boundgdminrhoh} \D&\ge& \frac{29}{12}-\frac{5 \sqrt{2}}{3}=0.05964 \mbox{ (up to five decimal places)}, \\\N&\le&\frac{47}{12}-\frac{8 \sqrt{2}}{3}=0.14543 \eqa
The operators corresponding to Eq. (\ref{optimalgdporj}) fail to be projectors and hence we resort to numerical techniques. Numerically, with $10^6$ randomly generated measurements, we have achieved the value $0.07966$ for GD and $0.143934$ for MIN. Thus, the bound for MIN is better than that of GD and both the measures are small. The measurement operators $\Pi_{1,2}=(1/2)(|0\ran\pm|1\ran)(\lan 0|\pm|\lan 1|), \Pi_{3,4}=(1/2)(|2\ran\pm|3\ran)(\lan 2|\pm\lan 3|)$ gives better result than the computational basis for MIN ($=0.142977$).

\section{Unextendible-product-basis (UPB) based bound entangled states}
\subsection{\label{bennettspyramid} Bennett \etal's $3\otimes3$ bound entangled state (Pyramid) \cite{BennettetalPRL99}} The first UPB based bound entangled state is given by
\bq\label{bennettetalsbespyramid} \rho_{Pyramid}=\frac{1}{4}\left(\mathbf{I}_3\otimes\mathbf{I}_3-\sum_{i=0}^4|\psi_i\ran\lan\psi_i|\right)\eq where
$|\psi_i\ran=|v_i\ran|v_{2i\mbox{ mod }5}\ran$; $v_i=N(\cos\frac{2\pi i}{5},\sin\frac{2\pi i}{5},h)$, $N=2/(5+\sqrt{5})^{1/2}$, $h=(1+\sqrt{5})^{1/2}/2$.

Using Eq. (\ref{lowerboundgd}) we have the following bound on GD \bqs\D\ge\frac{1}{32} \left(19-7 \sqrt{5}\right) = 0.10461\eqs
It can be easily seen that the operators in Eq. (\ref{optimalgdporj}) fail to be projectors, leading to difficulties in analytic calculations. However we note that these operators originated from a particular choice of unitary  and there may be other measurements (corresponding to different choices of unitaries) which would saturate the bound \cite{RanaParasharPRA12}. Indeed it turns out that the projection operators $\Pi_k=|p_k\ran\lan p_k|,k=1,2$ with \bqs |p_1\ran=\frac{1}{\sqrt{6}}\left(\sqrt{3},1,\sqrt{2}\right),\quad |p_2\ran=\frac{1}{\sqrt{3}}\left(0,-\sqrt{2},1\right)\eqs saturate this bound and hence the exact GD is given by \bq\label{finalgdpyramid}\D=\frac{1}{32} \left(19-7 \sqrt{5}\right) = 0.10461 \eq

Unfortunately, $\rho^A$ is degenerate and hence using the bound of Eq. (\ref{upperboundmin}), we get \bqs\N\le\frac{5\sqrt{5}}{48} \left(3-\sqrt{5}\right)= 0.17794\eqs As in the case of GD, this bound is not saturated by the operators in Eq. (\ref{optimalgdporj}). But with a careful analysis, we can get the analytic MIN in this case. First we note that the non degenerate eigenvector of $\rho^A$ is $|2\ran\lan2|$. In order to keep $\rho^A$ invariant, the other two projectors must be from span$\{|0\ran,|1\ran\}$ only  and so, without loss of generality, can be taken as $\Pi_{1,2}=(1/2)(I_2\pm x.\sigma)$ with $x=(a,b,c)\in \mathbb{R}^3, |x|=1$. Note that $\Pi_{1,2}$ act on 3-dimensional space, so their third row and column are entirely zeros. After some algebraic simplification, this measurement gives \bqs \|\rho-\Pi^A(\rho)\|^2=\frac{1}{8}\left(\sqrt{5}-2\right)\left(3+b^2\right)\eqs Clearly the maximum occurs at $b=1$, thereby we get the exact MIN as \bq\label{minpyramid}\N=\frac{3}{4} \left(\sqrt{5}-2\right)= 0.17705\eq and the optimal von Neumann projectors are given by \bqs\Pi_{1,2}=\frac{1}{2}(I_2\pm\sigma_y),\quad\Pi_3=|2\ran\lan2|.\eqs 

\subsection{Bennett \etal's $3\otimes3$ bound entangled state (Tiles) \cite{BennettetalPRL99}} Another UPB-based bound entangled state from the same work is
\bq\label{bennettetalsbestiles} \rho_{Tiles}=\frac{1}{4}\left(\mathbf{I}_3\otimes\mathbf{I}_3-\sum_{i=0}^4|\psi_i\ran\lan\psi_i|\right)\eq where
$|\psi_0\ran=|0\ran(|0\ran-|1\ran)/\sqrt{2}$, $|\psi_1\ran=(|0\ran-|1\ran)|2\ran/\sqrt{2}$, $|\psi_2\ran=|2\ran(|1\ran-|2\ran)/\sqrt{2}$,
$|\psi_3\ran=(|1\ran-|2\ran)|0\ran/\sqrt{2}$, $|\psi_4\ran=(|0\ran+|1\ran+|2\ran)(|0\ran+|1\ran+|2\ran)/3$.

\begin{figure}[ht]
\includegraphics[height=5cm]{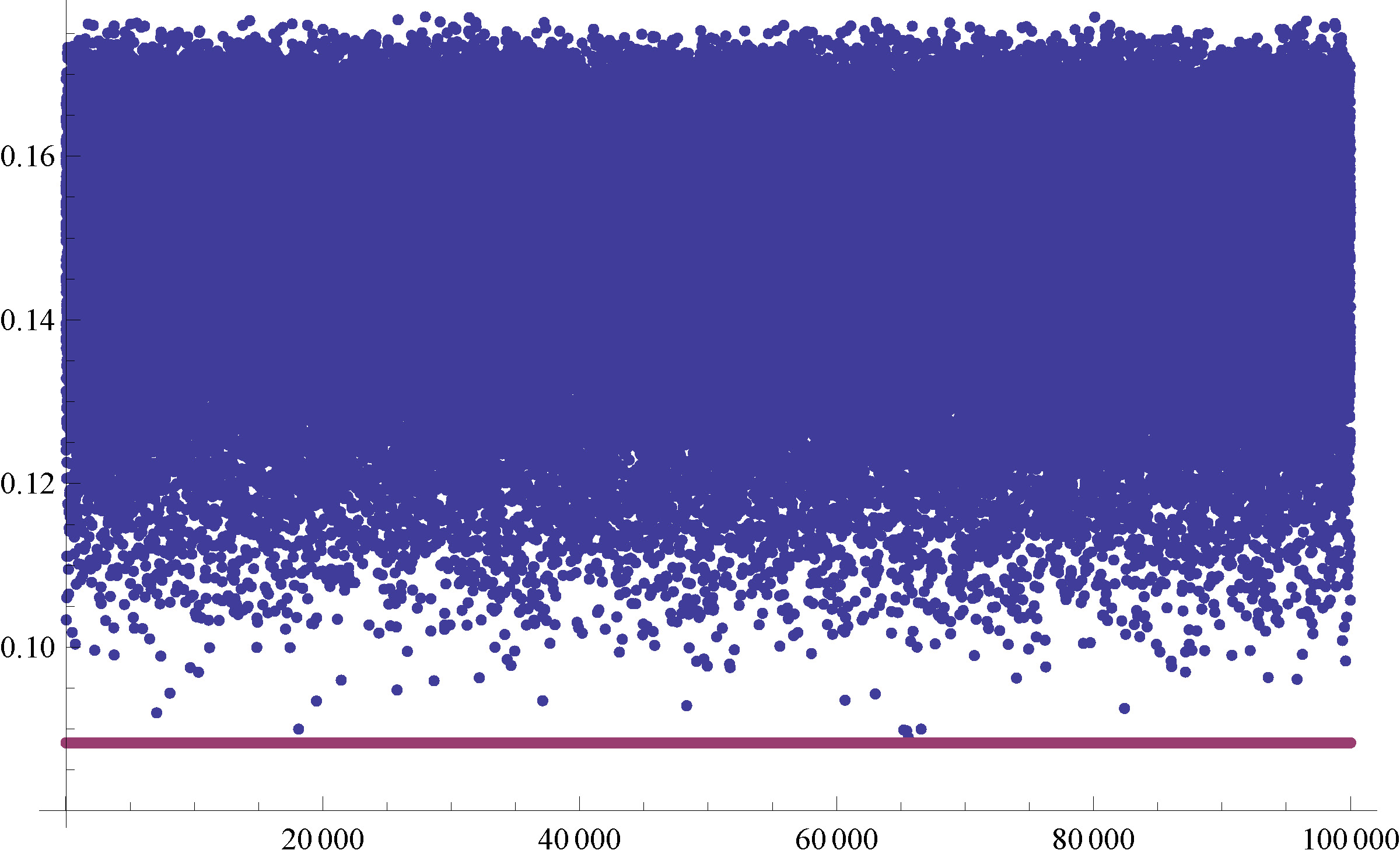}
  \caption{(Color online) A simulation of $10^5$ random measurements on the state in Eq. (\ref{bennettetalsbestiles}). The red solid line represents the bound of $\D\ge0.08832$, whereas we have achieved the value $0.088990$.}\label{tilesplot}
\end{figure}

For this state, analytic evaluation of GD is very complicated. Calculating even the bound in Eq. (\ref{lowerboundgd}) is quite difficult. For example, the eigenvalues of $G$ are $(19+2r\cos\theta)/64, 27/64$ (multiplicity $2$), $(19-2r\cos(\theta\pm \pi/3))/64$ and $0$ (multiplicity $3$), where $r=\sqrt{55}$ and $\theta=(1/3)\arctan(9\sqrt{1319}/244)$. Thus we have the following bound \bqas \D&\ge& \frac{1}{576} \left(65-2 \sqrt{55} \cos\left[\frac{1}{3} \arctan\left(\frac{9 \sqrt{1319}}{244}\right)\right]\right)\\&=&0.08832 \eqas We use numerical techniques to judge this bound. A simulation of $10^5$ measurements has been shown in FIG-\ref{tilesplot}. Numerically we have achieved the value $0.088990$. We note that the bound is quite good as it gives correct result up to two decimal places.

Fortunately, $\rho^A$ is non degenerate. Hence we have the exact value of MIN. The (unnormalized) eigen projectors of $\rho^A$ are  $(1,(5-\sqrt{33})/2,1)$, $(-1,0,1)$ and $(1,(5-\sqrt{33})/2,1)$. Since they correspond to the optimal measurement, we have \bq\label{mintiles} \N=\frac{95}{704} =  0.13494\eq

\section{Benatti \etal's $4\otimes 4$ bound entangled state \cite{BenattietalPLA04}} The state $\rho$ has the following matrix form in computational basis (the zero entries are not written to retain clarity of the nice pattern) \bqs\rho=\frac{1}{24}\left(
\begin{array}{cccccccccccccccc}
 1 & &  &  &  & -1 &  &  & &  & -1 &  &  &  &  & 1 \\
  & 3 &  &  & -1 &  &  &  &  &  &  & -1 & &  & -1 &  \\
  &  & 1 &  &  &  &  & -1 & -1 &  &  &  &  & 1 &  &  \\
  &  &  & 1 &  &  & 1 &  &  & 1 &  &  & 1 & &  &  \\
  & -1 &  &  & 3 &  &  &  &  &  &  & -1 &  &  & -1 &  \\
 -1 &  &  &  &  & 1 &  &  &  &  & 1 &  &  &  &  & -1 \\
  &  &  & 1 &  &  & 1 &  &  & 1 &  &  & 1 &  &  &  \\
  &  & -1 &  &  &  &  & 1 & 1 &  &  &  &  & -1 &  &  \\
  &  & -1 &  &  &  &  & 1 & 1 &  &  &  &  & -1 &  &  \\
  &  &  & 1 & &  & 1 &  &  & 1 &  &  & 1 &  &  &  \\
 -1 &  &  &  &  & 1 &  &  &  &  & 1 &  &  &  &  & -1 \\
 & -1 &  &  & -1 &  &  &  & &  &  & 3 &  &  & -1 &  \\
  &  &  & 1 &  &  & 1 &  &  & 1 &  &  & 1 &  &  &  \\
  &  & 1 &  &  &  &  & -1 & -1 &  &  &  &  & 1 &  &  \\
  & -1 &  &  & -1 &  &  & &  &  &  & -1 &  &  & 3 &  \\
 1 &  &  &  &  & -1 &  & &  &  & -1 &  &  &  &  & 1
\end{array}
\right)\eqs This state arises in the study of non decomposable positive maps and it turns out that it has some similarities with the Werner and isotropic states. For example, it has maximally mixed subsystems and, as we are going to show, the amount of GD coincides with the amount of MIN and the optimal measurement is in the computational basis.

To evaluate GD, we note that $x=0$ and $TT^t=(4/9)I_{15}$. The corresponding operators of Eq. (\ref{optimalgdporj}) are the computational basis $\{|k\ran\lan k|\}$ and hence the exact value of GD turns out to be $\D=1/9$. As $\rho^A$ is fully degenerate, finding MIN in the general case is very difficult. However, we note that if $x=0$ and the eigenvalues of $TT^t$ are equal, the two bounds in Eq. (\ref{lowerboundgd}) and Eq. (\ref{upperboundmin}) coincide and hence if one of them is saturated, the other follows immediately, with the same measurement (this was exactly the case in $4\otimes4$ Werner and isotropic states \cite{RanaParasharPRA12}). Thus we have $\D=\N=1/9$.

A quite surprising property of this state is that it becomes separable, if we view it (the matrix representation) as the density matrix of a $2\otimes 8$ system (in general, the same matrix may represent a separable or entangled system depending on the dimensions of the subsystems involved \cite{KCHaPRA11}. We also note that based on some incorrect derivation, the author of Ref. \cite{WChengCEC08} has shown that the state will remain bound entangled, which is wrong). By a tedious but somewhat straightforward calculation it follows that $x=0$ (though $y\ne0$) and $TT^t=(4/9)I_3$. Hence for this $2\otimes8$ case, $\D=\N=1/9$ holds too. Out of curiosity, we have checked that the same holds for the $4\otimes4$ Werner and isotropic states, though they are not bound entangled. The $m\otimes m$ Werner and isotropic states are respectively given by \bqa\rho_w&=&\frac{m-z}{m^3-m}\mathbf{I}+\frac{mz-1}{m^3-m}F,\quad z\in[-1,1]\label{wernerstatedef}\\ \rho_i&=&\frac{1-z}{m^2-1}\mathbf{I}+\frac{m^2z-1}{m^2-1}|\Psi\ran\lan\Psi|,\quad z\in[0,1]\label{isotropicstatedef} \eqa
where $F=\sum|kl\ran\lan lk|$ and $|\Psi\ran=(1/\sqrt{m})\sum|kk\ran$. It is well known \cite{LuoFuPRA10,RanaParasharPRA12} that \bqa\D({\rho}_w)=\N({\rho}_w)&=& \left(\frac{mz-1}{m^2-1}\right)^2, \label{wernergd44}\\\D({\rho}_i)=\N({\rho}_i)&=&\left(\frac{m^2z-1}{m^2-1}\right)^2 \label{isotropicgd44} \eqa
The $2\otimes 8$ dimensional state which has the same matrix representation (in computational basis) as the $4\otimes 4$ dimensional $\rho_w$, has $x=0$ (though $y\ne 0$) and $TT^t=16(1-4z)^2I_3/225$. Hence this $2\otimes 8$ state has the same $\D$($=\N$) as the $4\otimes 4$ Werner state. The same also holds for the $4\otimes 4$ dimensional $\rho_i$.

Indeed, $\D(=\N)$ remains same for the $2m\otimes 2m$ Werner (and isotropic) states even when viewed as $2\otimes 2m^2$ dimensional systems. However, the amount of entanglement, as quantified by negativity, may change \cite{RanaParasharPRAR12}.

\section{Discussion and conclusion}
The bound entangled states appeared more than a decade ago and have been studied extensively from several perspectives. The notion of geometric discord and measurement induced nonlocality has been introduced very recently and since then these measures have been evaluated for many classes of states. In this work for the first time we have evaluated these measures for bound entangled states. We found that both the measures attain relatively very small value.  Besides the states presented here, we have studied some other bound entangled states which also give small values. It would be interesting to find bound entangled states having GD and MIN close to one. Indeed, it is a challenging problem to find the maximum value of these measures that the bound entangled states can attain. We hope this work will generate interest to explore these issues further.

As mentioned earlier, the evaluation of measurement induced nonlocality becomes very difficult when $\rho^A$ has degenerate spectrum. In Sec. \ref{bennettspyramid}, we have shown how the case of small degeneracy could be tackled. Recently, a general approach to deal with the degenerate cases has been outlined in \cite{MirafzalietalAR11}.

To conclude, we have evaluated two post-entanglement measures of quantumness, namely geometric discord and measurement induced nonlocality for various bound entangled states. We have given analytic results in most of the cases. In particular, duality between the two measures has been reflected well in the Horodecki's state $\rho_{\beta}$ given in Eq. (\ref{allhbes}). When viewed as $2\otimes 8$ system, similar to the Werner and isotropic states, not only the notion but also the amount of quantumness of Benatti \etal's $4\otimes 4$ bound entangled state remains the same.

\end{document}